\documentclass{eage2012}

\usepackage[active]{srcltx}  
\usepackage{graphicx}

\begin{document}

\title{Automated seismic-to-well ties?}
\author{Roberto Henry Herrera and Mirko van der Baan}
\maketitle

\begin{abstract}
The quality of seismic-to-well tie is commonly quantified using the classical Pearson's correlation coefficient. However the seismic wavelet is time-variant, well logging and upscaling is only approximate, and the correlation coefficient does not follow this nonlinear behavior. We introduce the Dynamic Time Warping (DTW) to automate the tying process, accounting for frequency and time variance. The Dynamic Time Warping method can follow the nonlinear behavior better than the commonly used correlation coefficient. Furthermore, the quality of the similarity value does not depend on the selected correlating window. We compare the developed method with the manual seismic-to-well tie in a benchmark case study.


\end{abstract}

\section{Introduction}

Well logs are commonly used as ground truth to correlate the seismic signal with the earth's stratigraphy \citep{White2003}. In this process, a wavelet is first estimated from the seismic trace, and then it is convolved with the reflectivity calculated from the well logs (sonic log and bulk density log). Iterative techniques are used to estimate the wavelet with correct phase and amplitude spectrum by matching the actual seismic trace at the well position \citep{Hampson-Russell1999}. Tying the seismic traces to the well logs aims to minimize the differences in the way seismic data and well logs measure the same parameters \citep{Burch2002}, but with different resolution.

The quality of the tie between the synthetic and the seismic trace is based on the correlation coefficient, which is limited to linear features. The time-variant nature of the seismic wavelet adds nonlinearities to the trace which cannot be followed by a linear metric. Thus, wavelet phase mismatches frequently occur between the final processed seismic data and the synthetic seismograms created from well logs. This fact leads to potential complications in stratigraphic and structural interpretation \citep{VanderBaan2008}. We propose a new method to match these time series accounting for frequency and time variance. We argue that Dynamic Time Warping (DTW) method can follow these changes and furthermore the quality of fit is not limited to the selected correlation window.

Nonlinearities in time series representing physical processes are common in many areas (speech processing \citep{Rabiner1993Fundamentals}, medicine, industry and finance \citep{Keogh02exactindexing}). More recently, these concepts have been improved by \cite{Keogh2003} and \cite{Keogh2004}. DTW is a robust tool to match time series even if they are out of phase or time shifted \citep{Keogh02exactindexing}.

Related works using DTW in seismic applications come with the attempts to automate well-to-well log correlation \citep{Lineman1987a, Steven2004}. Well logs from different wells are correlated to infer common earth features. \cite{Steven2004} found that the cross-correlation was unable to follow local distortions such as stretching or shrinking of stratigraphic intervals, typical of logs collected even from closely spaced wells. \cite{Anderson1983} seek for correspondence between features in the logs of various wells using dynamic programming tools. To our knowledge, there are no previous reports of dynamic programming applied to the seismic-to-well tie problem.

\section{Method and Theory}


The correlation coefficient is commonly used to measure the quality of the seismic-to-well tie \citep{Hampson-Russell1999}. Comparing two (time-dependent) sequences $S :=[s_1,s_2, \ldots,s_n]$ and $T :=[t_1,t_2, \ldots,t_n]$, both of length $n$, will give a correlation coefficient at the time lag $\tau$:
\begin{equation}\label{eq1:ccorr}
\gamma_{ST}(\tau) = \frac{\sum_{i=1}^n [S(i)-\mu_S][T(i+\tau)-\mu_T]}{(\sum_{i=1}^n [S(i)-\mu_S]^2 \sum_{i=1}^n [T(i)-\mu_T]^2)^{1/2}},
\end{equation}
\noindent where $\mu_x$ is the average of trace $x$.

This measure works well if a constant time shift $\tau$ characterizes both signal. But the majority of geophysical applications have time alignment problems  \citep{Anderson1983}. When this time alignment is constant, the problem is reduced to the correction of the time lag by cross correlation. But this measure fails to find the best matching in nonlinear cases.

An alternative to the cross correlation is to find the Euclidean distance ($L_2-$norm) between the two time series \citep{Keogh2003}:
\vspace{-1mm}
\begin{equation}\label{eq1:euc}
    D_{euclid}(S,T) = \sqrt{\sum_{i=1}^n (S(i)-T(i))^2}.
\end{equation}
\vspace{-1mm}
\noindent where $D_{euclid}(S,T)$ is the one-to-one distance between the synthetic $S$ and the trace $T$.

The Euclidean distance ($L_2-$norm) is the most widely used distance measure. It is trivial to implement but also is very sensitive to small distortions in the time axis \citep{Keogh2003,BerndtC94}. Taking the advantages of the Euclidean distance and adapting it for nonlinear matching, \cite{BerndtC94} proposed the Dynamic Time Warping technique as we know it today.

\begin{figure}[h!]
  \begin{center}
   \includegraphics[width=.55\linewidth,angle=0]{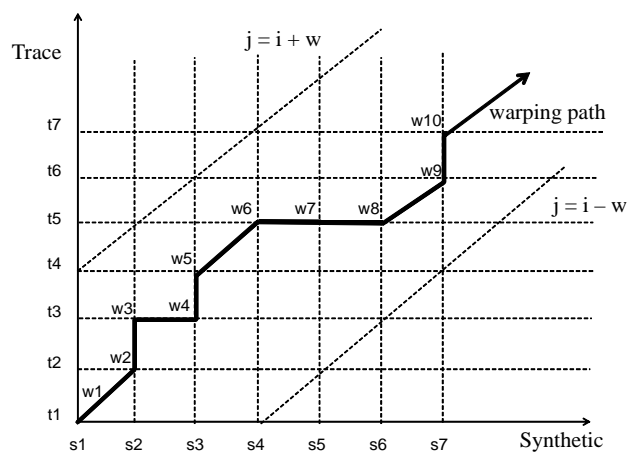}
  \end{center}
  \vspace{-4mm}
  \caption{Warping path at the minimum distance of two sequences. The warped versions of this example are $S_{warped} =s1,s2,s2,s3,s3,s4,s5,s6,s7,s7$ and $T_{warped}= t1,t2,t3,t4,t5,t5,t5,t6,t7$.}
\label{fig:warpingpath}
\end{figure}

DTW distance can accommodate stretching and squeezing in the time series by linear programming. It uses the Euclidean distance as the initial metric but allows for the one-to-many non-linear alignment. The warping distance is represented as the minimum path in a grid representation of both sequences. In Figure \ref{fig:warpingpath} the warping path, $W =w_1,w_2,\dots,w_k$, aligns the elements of $S$ and $T$ is such a way that the distance between them is minimized.

In this matrix the square distance in the elements ($i^{th},j^{th}$) is calculated by:
\begin{equation}\label{eq:distance}
    \delta(s_i,t_i)= (s_i - t_i)^2
\end{equation}
To find  the best alignment between these two sequences we have to retrieve the path through the matrix that minimizes the total cumulative distance between them \citep{Keogh02exactindexing} as illustrated in Figure \ref{fig:warpingpath}. The optimal path minimizes the total warping cost \citep{BerndtC94} is:
\begin{equation}\label{eq:warpingcost}
    DTW(S,T)= \min_W \sum_{k=1}^p \delta(w_k),
\end{equation}

\noindent where each $w_k$ corresponds to a point $(i,j)_k$. From Figure \ref{fig:warpingpath} we can extract the first samples of the new warped sequences as $S_{warped} =[s_1,s_2,s_2,s_3,s_4]$ and $T_{warped} =[t_1,t_2,t_3,t_3,t_4]$. In this way sequences are accelerated or decelerated along the time axis.
From the linear programming point of view the problem is to find the minimum cost warping path, $|i_k - j_k| \leq w_k$. The dynamic programming approach uses the following recurrence to find the warping path \citep{BerndtC94}:
\begin{equation}\label{eq:cumulative}
    \gamma(i,j)= \delta(s_i,t_j) + \min [\gamma(i-1,j),\gamma(i-1,j-1),\gamma(i,j-1)],
\end{equation}
\noindent where $\delta(s_i,t_j)$ is the distance defined in (\ref{eq:distance}), and the cumulative distance $\gamma(i,j)$ is the sum distance between the current elements and the minimum cumulative distance of the three neighboring cells.

We apply the DTW algorithm to obtain a first seismic-to-well tie alignment between observed seismic data and the synthetic trace created from the well logs.

\section{Results}

The dataset used in our experiments consist of a 3D post-stack seismic profile, with 13 wells and their correspondent logs \citep{Hampson-Russell1999}. We use well 08-08 to estimate the wavelet which is subsequently used in all ties. The original observed data and unstretched synthetics are subjected to the DTW approach.

\begin{figure}[h]
  \begin{center}
   \includegraphics[width=.5\linewidth,angle=0]{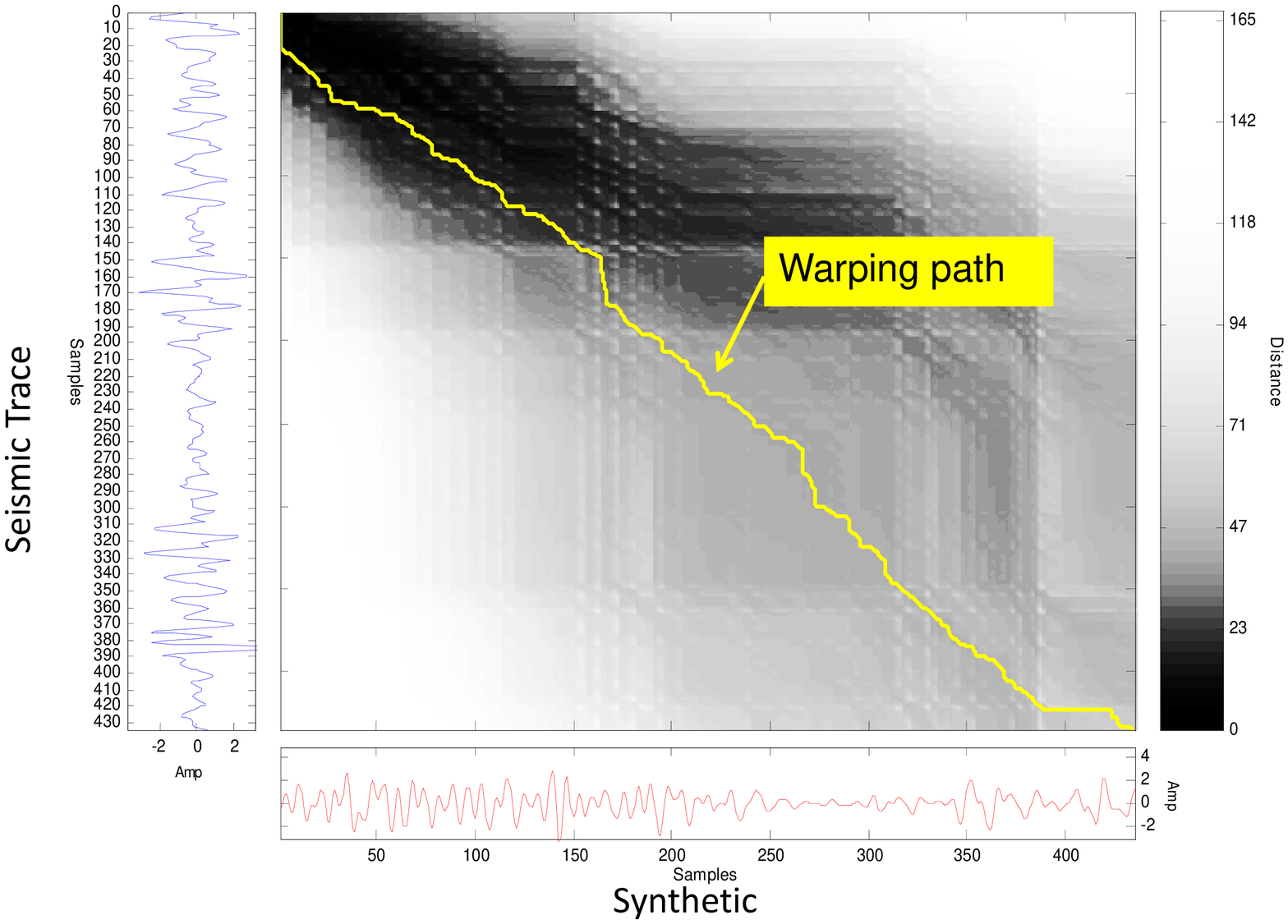}
   \includegraphics[width=.45\linewidth,angle=0]{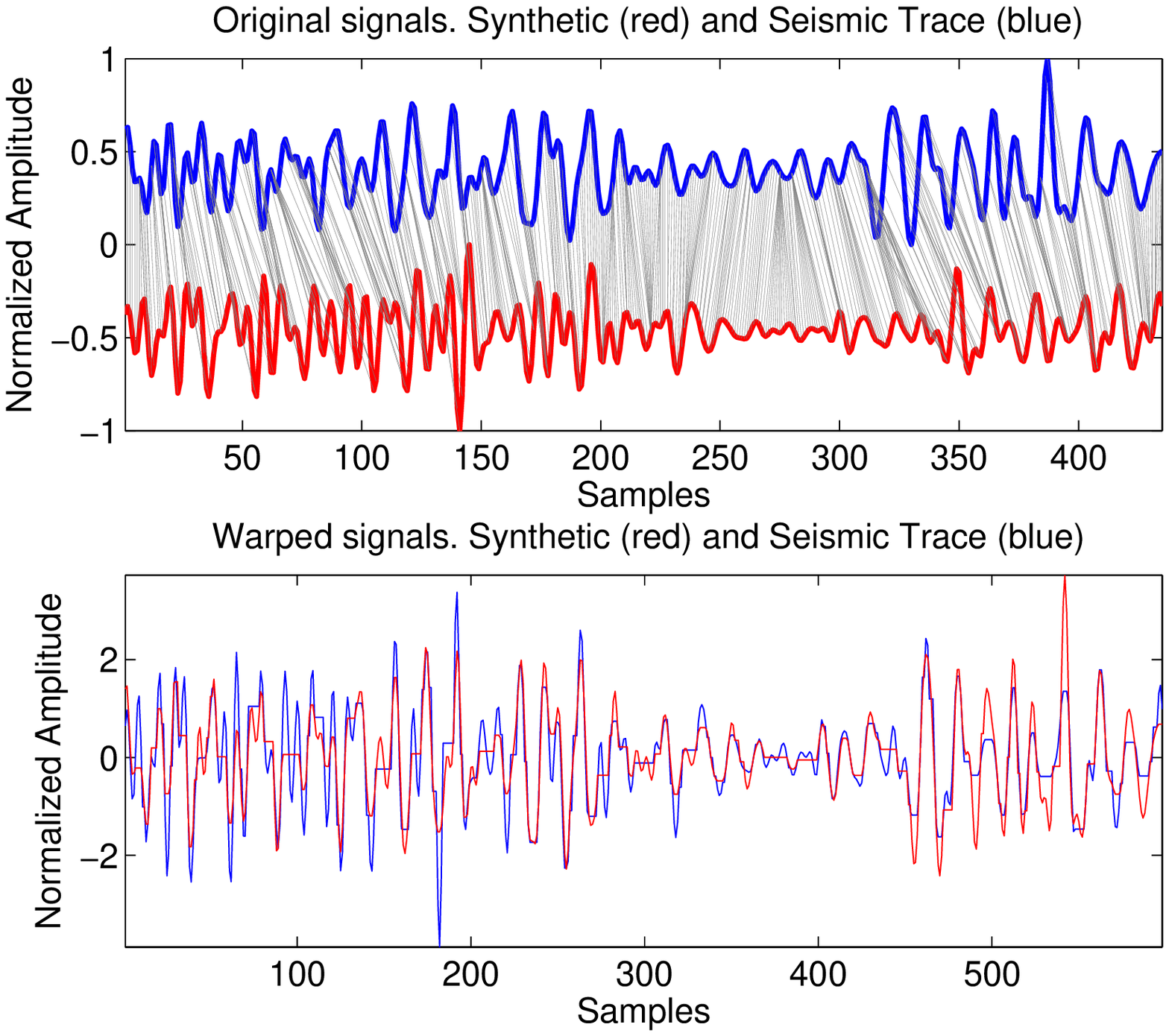}
  \end{center}
  \vspace{-4mm}
  \caption{Left plot shows the warping path represented by the yellow nonlinear trace, the lateral and bottom signals are the Trace and the Synthetic respectively. The right upper plot shows the original signals and their corresponding matching points, note that nonlinear and one-to-many associations occur in both ways. The right bottom plot depicts the warped version of both signals.}
\label{fig:wpwell01}
\vspace{-4mm}
\end{figure}

The estimated warping path for well 01-08 is shown in Figure \ref{fig:wpwell01} (left). Note the nonlinear relationship between both signals.

The effect of the DTW over the signals while finding the best match is shown in Figure \ref{fig:wpwell01} (right). Instead of having a one-to-one comparison between these two signals a nonlinear alignment between their matching points is found. By applying the time warping correction to the original signals we obtain their warped version as is shown in the lower right plot of Figure \ref{fig:wpwell01}. They are both stretched to match their common features.

\begin{figure}[h]
  \begin{center}
   \includegraphics[width=.7\linewidth,angle=0]{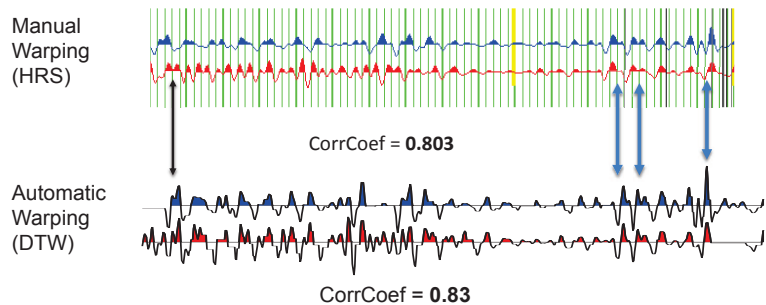}
  \end{center}
  \vspace{-40mm}
  \caption{Case 1: Well 01-08. Upper plot shows the manual seismic-to-well tie, the correlation of this tie was 80 \%. Bottom plot shows the automatic tie with a correlation of 83 \%. A good agreement is found between both the manual and automatic well tie.}
\label{fig:case1}
\vspace{-1mm}
\end{figure}

The correlation coefficients are estimated over the entire length of the well log for the DTW. Results of this measure are similar to the ones obtained in the time window from 800 ms to 1100 ms for the manual tie. Figure \ref{fig:case1} shows the results for the manual seismic-to-well tie for Case 1: well 01-08. The DTW method was able to match similar events along the full seismic trace and gives a similar correlation coefficient to the one obtained using the manual method.

\begin{figure}[h]
  \begin{center}
   \includegraphics[width=.75\linewidth,angle=0]{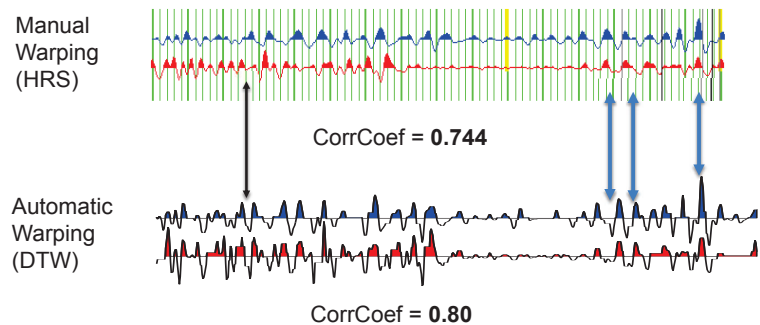}
  \end{center}
  \vspace{-38mm}
  \caption{Case 2: Well 16-08. Upper plot shows the manual seismic-to-well, the correlation of this tie was 74 \%. Bottom plot shows the automatic tie with a correlation of 80 \%.}
\label{fig:case2}
\vspace{-1mm}
\end{figure}

Figure \ref{fig:case2} shows a second example where a better correlation is obtained in the automated method than the manual result. Note the correspondence between both traces in the bottom plot.
\vspace{-3mm}

\section{Conclusions}

We have implemented dynamic time warping to automate the seismic-to-well tie procedure, as this approach provides an optimal solution for matching similar events. We strongly advocate however to beware fully automated and non-supervised applications of this method, as a visual quality control of the end result remains highly advisable.

Future efforts will be oriented to analyze the effect of the estimated wavelet over the resultant warped sequences.
Many other applications of DTW are envisionable for seismic data. These include log-to-log correlations, alignment of baseline and monitor surveys in 4D seismics, PP and PS wavefield registration for 3C data.

\section{Acknowledgements}

The authors thank Hampson-Russell for software licensing and the Sponsors of the Blind Identification of Seismic Signals (BLISS) project for their financial support.

%
%

\bibliographystyle{firstbreak}
\bibliography{s2welltie}

\end{document}